%% file: main.tex
\documentclass{article}
\usepackage{spconf,amsmath,graphicx}
\usepackage{amssymb}
\usepackage{booktabs}

\usepackage{enumitem}
\setlist{nosep, leftmargin=12pt}

\graphicspath{{fig/}}

\newcommand{\zclin}{z_{\mathrm{clin}}}
\newcommand{\aid}{a_{\mathrm{id}}}
\newcommand{\zid}{z_{\mathrm{id}}}
\newcommand{\Dmod}{\mathcal{D}_{\mathrm{mod}}}
\newcommand{\Drel}{\mathcal{D}_{\mathrm{rel}}}
\newcommand{\Dtest}{\mathcal{D}_{\mathrm{test}}}
\newcommand{\Dval}{\mathcal{D}_{\mathrm{mod,val}}}
\newcommand{\auccross}{\mathrm{AUC}_{\mathrm{cross}}}

\newcommand{\Lc}[1]{\mathcal{L}_{#1}}

\title{LAPIS: Clinically Constrained Patient-Identity Residual Learning for
       Chest X-Ray Anonymization}
\name{Nishi Kanta Paul$^{\star}$ \qquad Md Shihabul Islam Shovo$^{\star}$ \qquad D\'{e}sir\'{e} Sidib\'{e}$^{\dagger}$}
\address{$^{\star}$NIMISHES Lab, Dhaka, Bangladesh \qquad
         $^{\dagger}$IBISC, Evry Paris-Saclay University, France \\
         {\small nishikantapaul108@gmail.com, shihabul900@gmail.com, drodesire.sidibe@univ-evry.fr}}
\begin{document}
\ninept
\maketitle
\input{sections/1_abstract}
\input{sections/2_introduction}
\input{sections/3_related_work}
\input{sections/4_methodology}
\input{sections/5_experiments}
\input{sections/6_results}
\input{sections/7_conclusion}
\input{sections/8_backmatter}

\bibliographystyle{IEEEbib}
\bibliography{references}

\end{document}

%% file: sections/1_abstract.tex
\begin{abstract}
Biometric information in chest radiographs limits their release for research,
as patients can be re-identified. We propose
LAPIS, which jointly adapts a spatial chest-radiograph latent space so that
patient-discriminative information concentrates in a compact, replaceable channel residual while most clinically relevant content stays in the complement. The residual is defined by an exactly
semi-orthogonal basis, a spatial adversary suppresses identity leakage from the
complement, and replacement-aware training teaches the decoder cross-patient
residual swaps under output identity, clinical and realism constraints, so that
replacing the residual yields a valid image. At release, the residual is
replaced by a mixture of donor residuals. On patient-disjoint model, release and
test sets from ChestX-ray14, we evaluate attackers adapted to the anonymized
setting, a structural source-retrieval attack, and pathology classifiers
trained on released images and tested on real ones. LAPIS reduces
worst-case cross-study linkability from $0.961$ to $0.791$ AUC at a macro AUROC
cost of $0.012$ relative to reconstruction; a deformation-based baseline,
however, remains better at structural source retrieval.
\end{abstract}


%% file: sections/2_introduction.tex
\section{Introduction}
\label{sec:intro}

Large public chest radiograph collections are a standard source for training
thoracic diagnostic models, but sharing them is impeded because the images
are identifying: deep verification networks recover patient identity
in ChestX-ray14 with an ROC AUC approaching $0.99$~\cite{packhauser2022reid}, so
removing burned-in text and header metadata leaves substantial re-identification
risk. Anatomy, not annotation, encodes biometrics, so anonymization must change
image content.

\begin{figure*}[t]
\centering
\includegraphics[width=0.92\textwidth]{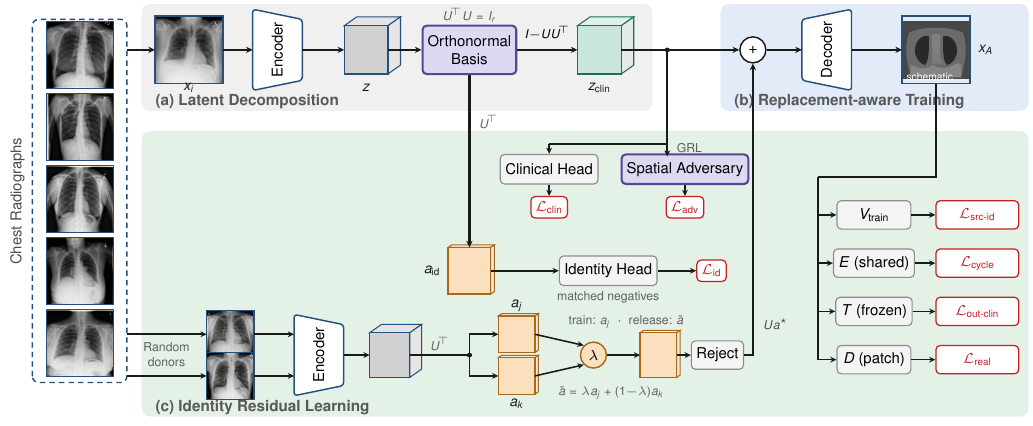}
\caption{LAPIS. (a) A semi-orthogonal basis $U$ ($U^{\!\top}U=I_r$), shared
across locations, splits the latent into a rank-$r$ patient-identity residual
$\aid$ and a clinical complement $\zclin$. (b) The decoder is trained on
cross-patient residual swaps under output identity, cycle, clinical and realism
constraints. (c) A metric loss with matched negatives makes $\aid$
patient-discriminative; a spatial adversary suppresses identity in $\zclin$. At
release, screened donor mixtures $\tilde{a}=\lambda a_j+(1-\lambda)a_k$ from
$\Dmod$ replace $\aid$. The output card is schematic, not a model prediction.}
\label{fig:arch}
\end{figure*}

Existing chest radiograph anonymizers (Sec.~\ref{sec:related}) leave three
gaps. First, privacy is typically measured with attackers never trained on the
anonymization process, so a drop in re-identification can reflect domain shift
rather than identity removal; separating the two requires a reconstruction
control. Second, linking an anonymized image to its own source and to another
study of the same patient are distinct risks, rarely reported separately.
Third, latent-manipulation anonymizers assume the decoder produces valid
images from recombined latents, which their training does not guarantee.

LAPIS closes these gaps without assuming that identity already resides in a
separable subspace; it learns one. Our hypothesis is narrow: not linear disentanglement of identity
and pathology, but the existence of a compact low-rank channel residual carrying
patient information whose replacement lowers adaptive linkability at a
detectable clinical cost. Our contributions are:

\begin{itemize}
  \item We formulate anonymization as patient-identity residual learning
  in the spatial latent, adapting the representation so that a compact channel
  subspace encodes transferable patient information while a spatial adversary
  suppresses leakage from the complement.
  \item We propose replacement-aware training, decoding cross-patient residual
  swaps under output identity, realism and clinical constraints, so the swap's
  image-level validity is trained, not assumed.
  \item We evaluate with attackers adapted to the anony\-mized domain and a
  structural source-linkability attack, training on released and testing on
  real images over patient-disjoint cohorts.
\end{itemize}

%% file: sections/3_related_work.tex
\section{Related work}
\label{sec:related}

Siamese verification networks trained on repeated studies show that chest
radiographs behave as a biometric~\cite{packhauser2022reid}, and recent models
generalize to multi-modal re-identification~\cite{tian2025medreid}; since such
attackers can learn from any released data, we treat attacker adaptivity as
intrinsic to the protocol. Anonymization follows two routes. PriCheXy-Net
learns a deformation field guided by an abnormality classifier and a
verification network, lowering re-identification from $0.818$ to $0.577$ AUC on
ChestX-ray14~\cite{packhauser2023prichexy}. Latent methods operate on codes:
anonymous radiographs are sampled from a latent diffusion model with a
privacy-enhancing sampling rule~\cite{packhauser2023ldm}, or each radiograph's
latent code is projected and optimized under paired identity and utility
losses~\cite{li2025gmia}, though per-image optimization is costly at release
scale and yields no reusable anonymizer. Disentanglement has anonymized
case-based explanations, keeping decision-relevant
properties~\cite{montenegro2024disentangle}, and face anonymization conditions
generation on another identity~\cite{maximov2020ciagan}. A radiograph, however,
must stay medically correct, so our clinical constraints act at latent and
output levels.

%% file: sections/4_methodology.tex
\section{Methodology}
\label{sec:method}

LAPIS anonymizes an image by replacing one component of its latent code. We
first fix the threat model, data and backbone (Sec.~\ref{sec:formulation}),
then learn a patient-identity residual separate from a clinical complement
(Sec.~\ref{sec:residual}), train the decoder to remain valid when that
residual is replaced (Sec.~\ref{sec:replace}), and construct the replacement
used at release (Sec.~\ref{sec:release}).

\subsection{Problem formulation}
\label{sec:formulation}

A data holder publishes anonymized images $x_A$ of patients unseen in training;
the attacker knows the technique and can retrain on the released data.
\emph{Same-image linkability} asks whether $x_A$ can be matched to its original
$x$, i.e.\ how much source-specific information survives. \emph{Cross-study
patient linkability} asks whether $x_A^{(1)}$ can be matched to a separate study
$x^{(2)}$ of the same patient, the stronger threat for repeatedly imaged
patients. Both are reported throughout.

Patients form three disjoint subsets. $\Dmod$ trains the anonymizer
and all auxiliary models, with $\Dval$ held out for hyperparameters and rank
selection; $\Drel$ is anonymized by the fixed model and forms the released
dataset on which privacy is measured; $\Dtest$ stays untouched and real, for
testing downstream pathology models.

\input{tables/main_comparison}

A convolutional autoencoder with downsampling factor $8$ maps a
$256\times256$ image $x$ through an encoder $E$ to a code $z=E(x)\in\mathbb{R}^{C\times H\times W}$
with $C=128$, $H=W=32$, and a decoder $G$ reconstructs $\hat{x}=G(z)$.
Training has two stages. Stage A pre-trains $E$ and $G$ alone with
$\Lc{rec}=\|x-\hat{x}\|_1+\lambda_s(1-\mathrm{SSIM}(x,\hat{x}))$~\cite{wang2004ssim}
and makes no privacy claims; Stage B starts from Stage A and trains the
full model of Secs.~\ref{sec:residual}--\ref{sec:replace}. Reconstruction
alone perturbs verification, so $G(E(x))$ is a control throughout.
Let $A_{\mathrm{orig}}$, $A_{\mathrm{rec}}$ and $A_{\mathrm{anon}}$ be the
macro AUROC on $\Dtest$ of a pathology model trained on original,
reconstructed and anonymized $\Drel$. This separates the autoencoder cost
$\Delta_{\mathrm{AE}}=A_{\mathrm{orig}}-A_{\mathrm{rec}}$ from the
anonymization-specific cost $\Delta_{\mathrm{anon}}=A_{\mathrm{rec}}-A_{\mathrm{anon}}$.
The goal is the lowest linkability subject to $\Delta_{\mathrm{anon}}\le\delta$,
with the margin $\delta=0.02$ fixed in advance.

\subsection{Patient-identity residual learning}
\label{sec:residual}

Stage B splits $z$ into a compact residual meant to carry patient identity
and a complement meant to carry clinical content (Fig.~\ref{fig:arch}a,c).
Let $U\in\mathbb{R}^{C\times r}$ be a channel basis satisfying $U^{\!\top}U=I_r$
exactly, via a standard Stiefel-manifold parametrization rather than a soft
penalty~\cite{lezcano2019orthogonal}. The same basis acts at every site $(h,w)$
(Fig.~\ref{fig:arch}a), giving a residual $\aid\in\mathbb{R}^{r\times H\times W}$
with $\aid(:,h,w)=U^{\!\top}z(:,h,w)$ and an exact orthogonal split
\begin{equation}
  \zid=U\aid,\qquad \zclin=(I-UU^{\!\top})z=z-\zid .
\end{equation}
A fixed basis with a spatially varying coefficient map is an inductive bias:
patient-specific appearance is expected to act through a few channel directions
of varying strength across the field of view. Stage B does not require such
structure in a frozen latent; it adapts $E$, $G$ and $U$ jointly to make it
useful.

Alongside $\Lc{rec}$, which Stage B retains, three losses shape this split.
A supervised contrastive loss~\cite{khosla2020supcon} $\Lc{id}$ on
$\aid$ makes the residual patient-discriminative, with positives from
different studies of the same patient. Negatives are matched on pathology
label profile, view position, pixel-spacing bucket and original dimension,
falling back to partial matches when no exact match exists, removing the
most obvious way to encode chronic disease or scan setup as a signature. A
spatial adversary $\Lc{adv}$, a small convolutional network, sees the
whole tensor $\zclin$ of a study pair and predicts same or different patient,
while a gradient-reversal layer~\cite{ganin2015grl} discourages the encoder from
leaking it; pooling $\zclin$ would obscure zero-mean spatial identity patterns.
A retention head $C_{\mathrm{clin}}$ predicts fourteen findings from $\zclin$
under $\Lc{clin}=\Lc{path}+\lambda_{KD}\Lc{KD}$, where $\Lc{path}$ is the binary
cross-entropy with the labels and $\Lc{KD}$ a distance $d$ between the
predictions of $C_{\mathrm{clin}}$ and those of an independent image-space
pathology classifier $T$ trained on $\Dmod$. Combined,
$\Lc{decomp}=\lambda_{rec}\Lc{rec}+\lambda_{id}\Lc{id}+\lambda_{adv}\Lc{adv}+\lambda_{clin}\Lc{clin}$.
The adversary is only a training regularizer; its accuracy is not a privacy
result.

\subsection{Replacement-aware training}
\label{sec:replace}

None of these steps ensures that $G(\zclin+U\tilde{a})$ is a valid radiograph
for a residual $\tilde{a}$ the decoder never saw with $\zclin$. Stage B
therefore also trains on replacements (Fig.~\ref{fig:arch}b): for a source $i$
and a random donor $j$ from the batch, the swap
$z_{i\leftarrow j}=\zclin^{\,i}+U\aid^{\,j}$ is decoded to
$x_{i\leftarrow j}=G(z_{i\leftarrow j})$, and four loss terms apply to it.
A patient verifier
$V_{\mathrm{train}}$, trained on $\Dmod$ and then frozen, is used only in
training and is disjoint from every evaluation attacker; it pushes the output
away from the source identity,
\begin{equation}
  \Lc{src\text{-}id}=\max\!\big(0,\ \mathrm{sim}(V_{\mathrm{train}}(x_i),
  V_{\mathrm{train}}(x_{i\leftarrow j}))-m_{src}\big),
\end{equation}
where $\mathrm{sim}$ compares $V_{\mathrm{train}}$ embeddings and $m_{src}$
is a margin. Re-encoding the swap ties both branches to their sources;
$\Lc{cycle}$ sums $\|(I-UU^{\!\top})E(x_{i\leftarrow j})-\zclin^{\,i}\|_1$ and
$\lambda_{c\text{-}id}\|U^{\!\top}E(x_{i\leftarrow j})-\aid^{\,j}\|_1$.
$T$ preserves the source clinical content at the output,
$\Lc{out\text{-}clin}=d\big(T(x_i),T(x_{i\leftarrow j})\big)$, and a
PatchGAN discriminator~\cite{isola2017pix2pix} provides $\Lc{real}$ to keep
recombined latents close to the radiograph manifold. The final Stage B loss is
\begin{equation}
  \begin{split}
  \Lc{} = \Lc{decomp}&+\lambda_{s\text{-}id}\Lc{src\text{-}id}
        +\lambda_{cyc}\Lc{cycle}\\
        &+\lambda_{o}\Lc{out\text{-}clin}+\lambda_{r}\Lc{real}.
  \end{split}
\end{equation}

\subsection{Release-time anonymization}
\label{sec:release}

After Stage B, $E$, $G$ and $U$ are frozen and applied to $\Drel$.
Copying one real patient's residual would substitute one identity for another.
Instead we take two donors $j,k$ from $\Dmod$ matched on coarse clinical and
acquisition properties, sample $\lambda\sim\mathrm{Beta}(\alpha,\alpha)$ and
construct (Fig.~\ref{fig:arch}c)
\begin{equation}
  \tilde{a}=\lambda a_j+(1-\lambda)a_k,\qquad x_A=G(\zclin+U\tilde{a}).
\end{equation}
$V_{\mathrm{train}}$ rejects and resamples any candidate whose similarity
to the source or to either donor exceeds a threshold fixed on $\Dval$.
Privacy is evaluated by separate attackers, never by this filter.

%% file: tables/main_comparison.tex
\begin{table*}[t]
\centering
\caption{Privacy on $\mathcal{D}_{\mathrm{rel}}$ under adaptive attackers; utility: macro AUROC / mAP on real $\mathcal{D}_{\mathrm{test}}$ of a pathology model trained on released images (mean $\pm$ s.d., 3 seeds). $\Delta_k^{\max}$: largest $\Delta_k$ over adequately powered findings. Bold: best anonymizer per column.}
\label{tab:main}
\setlength{\tabcolsep}{4.5pt}
\small
\begin{tabular}{l cc c cc c}
\hline
& \multicolumn{2}{c}{Learned linkability} & Structural & \multicolumn{2}{c}{Pathology utility} & Per-finding \\
\cline{2-3}\cline{5-6}
Method & AUC$_{\mathrm{cross}}\downarrow$ & AUC$_{\mathrm{same}}\downarrow$ & Rank-1 (\%)$\downarrow$ & AUROC$\uparrow$ & mAP$\uparrow$ & $\Delta_k^{\max}\downarrow$ \\
\hline
Original & 0.974 & 0.999 & 100.0 & 0.812\,$\pm$\,0.003 & 0.331 & -- \\
Reconstruction & 0.961 & 0.983 & 94.3 & 0.798\,$\pm$\,0.004 & 0.318 & -- \\
\multicolumn{7}{l}{\textit{Anonymization baselines}} \\
Full-latent noise & 0.897 & 0.912 & 71.8 & 0.741\,$\pm$\,0.006 & 0.256 & 0.081 \\
Random-$U$ replacement & 0.912 & 0.936 & 78.6 & 0.771\,$\pm$\,0.005 & 0.283 & 0.047 \\
Direct donor swap & 0.848 & 0.881 & 47.2 & 0.773\,$\pm$\,0.005 & 0.288 & 0.043 \\
PriCheXy-Net & 0.826 & \textbf{0.784} & \textbf{22.4} & 0.768\,$\pm$\,0.006 & 0.281 & 0.052 \\
GMIA-style & 0.869 & 0.904 & 58.1 & 0.779\,$\pm$\,0.005 & 0.294 & 0.038 \\
\textbf{LAPIS} & \textbf{0.791} & 0.868 & 34.7 & \textbf{0.786\,$\pm$\,0.004} & \textbf{0.304} & \textbf{0.034} \\
\hline
\end{tabular}
\end{table*}

%% file: sections/5_experiments.tex
\section{Experiments and results}
\label{sec:exp}

\subsection{Experimental setup}
\label{sec:setup}

\noindent\textbf{Data and cohorts.}
We use ChestX-ray14~\cite{wang2017chestxray}: $112{,}120$ frontal radiographs of
$30{,}805$ patients with fourteen finding labels, view position, original
dimensions and pixel spacing. Patients are disjointly partitioned into
$\Dmod$ ($18{,}510$ patients, $67{,}172$ images; $10\%$ withheld as $\Dval$),
$\Drel$ ($7{,}588$, $27{,}962$) and $\Dtest$ ($4{,}707$, $16{,}986$), which are
closely balanced: $43.1$--$43.3\%$ of patients have repeated studies and
$39.8$--$40.3\%$ of images are AP. Patients with repeated studies have a finding
in $50.0\%$ of images vs.\ $26.7\%$ for single-study patients, so patients
exposed to cross-study linkage are also sicker; reporting both threats on the
same release cohort retains this.

\noindent\textbf{Attackers.}
Each anonymizer is fixed and applied to $\Dmod$, and attackers are trained on
original--original, original--anonymized and anonymized--anonymized pairs from
this cohort, with thresholds tuned on $\Dval$ and evaluation on unseen $\Drel$
patients. The evaluation is thus adaptive: the attacker has already seen the
anonymized domain, so domain shift alone earns no score. We use ResNet-50,
ConvNeXt-T~\cite{liu2022convnext} and ViT-S~\cite{dosovitskiy2021vit} verifiers
and report the worst case. Pairs are capped at $100$ positive and $100$ matched
negative per patient so heavily followed-up patients cannot dominate, and
confidence intervals use a patient-clustered bootstrap. We also run a
non-learned structural attack retrieving the source by SSIM and normalized
cross-correlation, reported as Rank-1, and a secondary analysis with
acquisition-matched negatives.

\begin{figure*}[t]
\centering
\includegraphics[width=\textwidth]{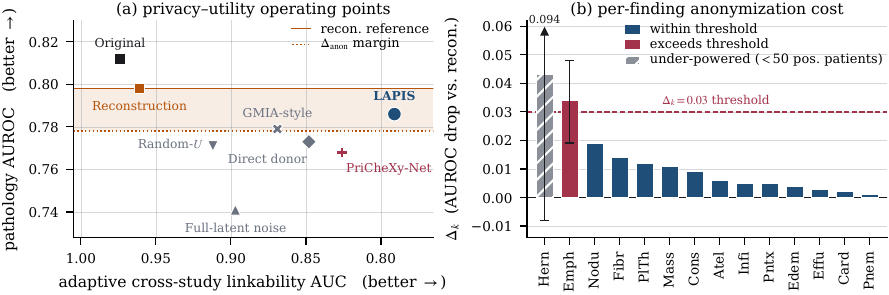}
\caption{(a) Operating points; only LAPIS and GMIA-style lie within the $0.02$
margin. (b) Per-finding cost with paired bootstrap intervals.}
\label{fig:pu}
\end{figure*}

\noindent\textbf{Utility.}
A DenseNet-121~\cite{huang2017densenet} is trained on original, reconstructed or
anonymized $\Drel$ and evaluated on real $\Dtest$ over three seeds, reporting
macro AUROC, mAP and the per-finding anonymization-specific cost $\Delta_k$
relative to reconstruction. By a pre-specified rule, a finding is materially
degraded if $\Delta_k>0.03$ and its paired bootstrap interval excludes zero,
assessed only for findings with at least $50$ positive test patients; Hernia
($28$) is high-uncertainty.

\noindent\textbf{Baselines.}
Besides the original and reconstruction references, we assess full-latent
Gaussian perturbation with noise scale set on $\Dval$, replacement with a random
orthonormal $U$ of equal rank, direct one-to-one swapping with another patient,
and PriCheXy-Net~\cite{packhauser2023prichexy}. We also add a GMIA-style latent
projection with per-image optimization~\cite{li2025gmia}; designed for
MIMIC-CXR, its ChestX-ray14 version is an adaptation, not a replication. All
baselines share the release cohort, attackers, pathology model and bootstrap
protocol.

\noindent\textbf{Implementation.}
Stage A takes $60$ epochs and Stage B $40$, with AdamW at $10^{-4}$ and $5\times10^{-5}$ respectively, batch size $32$, on two NVIDIA A100 40GB GPUs.
Loss weights and the mixing concentration $\alpha$ are chosen on $\Dval$, never
on the test set. The rank is selected by output-level validation, not by the
directly supervised latent head: from $r\in\{4,8,16\}$ we choose the
$r^{\star}$ minimizing worst-attacker cross-study AUC subject to $\Delta_{\mathrm{anon}}\le\delta$ on $\Dval$.
There, $r=16$ violates the constraint ($0.774$ vs.\ a $0.781$ floor) and $r=8$
gives the lowest admissible linkability ($0.803$ vs.\ $0.847$ for $r=4$), so
$r^{\star}=8$, i.e.\ $6.25\%$ of latent channels.

%% file: sections/6_results.tex
\subsection{Results and discussion}
\label{sec:results}

\noindent\textbf{Privacy.}
Reconstruction alone lowers worst-attacker cross-study linkability only from
$0.974$ to $0.961$ (Table~\ref{tab:main}): an autoencoder pass is not
anonymization, and its distribution shift fools no attacker acquainted with the
anonymized domain. Against this reference LAPIS reaches $\auccross=0.791$
($95\%$ CI $[0.778,0.804]$), beating geometry-sensitive PriCheXy-Net, $0.826$
($[0.812,0.840]$, non-overlapping), and full-latent isotropic noise, $0.897$,
the least efficient of the three per unit of utility lost, which again suggests
that identity lies in the structure of the latent. Our adaptive assessment of
PriCheXy-Net exceeds its originally reported
$0.577$~\cite{packhauser2023prichexy} because we measure cross-study
linkability, not source verification; hence all comparisons use one attacker
protocol, not cited numbers.

\noindent\textbf{Where the method is weaker.}
The structural attack identifies the source in $22.4\%$ of PriCheXy-Net
samples against $34.7\%$ for LAPIS, and PriCheXy-Net also wins on same-image
source verification ($0.784$ vs.\ $0.868$). This follows from our inductive
bias: a shared channel basis alters appearance but largely preserves the
anatomy's global geometry, kept for clinical reasons, whereas a learned
deformation attacks it directly. As identity has a geometric component,
releases needing robust same-image unlinkability benefit from a
geometry-sensitive residual; LAPIS is optimized for the cross-study threat and
pays for it here. Both stay far below the reconstruction control's $94.3\%$,
so LAPIS does substantially change source-specific structure, just less than
deformation.

\begin{figure}[t]
\centering
\includegraphics[width=0.93\columnwidth]{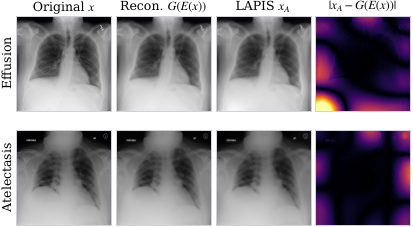}
\caption{Released images preserve lung-field appearance; the
anonymization-specific change is smooth and peripheral.}
\label{fig:qual}
\end{figure}

\noindent\textbf{Utility.}
Training on released anonymized images costs $0.012$ macro AUROC relative to
reconstruction, inside the margin $\delta=0.02$ fixed a priori
(Fig.~\ref{fig:pu}a). GMIA-style also clears it, at $0.019$, but reaches only
$0.869$ linkability against $0.791$ for LAPIS. The autoencoder pass costs a
similar $0.014$, so the representation bottleneck costs about as much as the
anonymizer built on it. Most drops are small (Fig.~\ref{fig:pu}b), but
Emphysema drops by $0.034$ with a paired interval of $[0.019,0.048]$, excluding
zero, and is thus materially degraded under our criteria. It is a diffuse,
low-contrast texture finding, and the smooth low-frequency modulation of
residual replacement (Fig.~\ref{fig:qual}) may interfere with exactly that cue.
Hernia drops by $0.043$, but with $28$ positive test patients its interval
includes zero. One degraded finding of thirteen adequately powered ones is
isolated, but a real cost to weigh against a release's purpose.

\input{tables/ablation}

\noindent\textbf{Ablations.}
Removing replacement-aware training hurts most, damaging privacy and utility
jointly ($0.884$, $0.742$; Table~\ref{tab:ablation}): without it the decoder
cannot handle the mismatched residual, so substitution neither displaces
identity nor yields acceptable images. A pooled instead of spatial adversary
leaves utility essentially unchanged ($0.784$), as the adversary constrains only
leakage through $\zclin$, but loses $0.058$ AUC of privacy, supporting the view
that pooling hides zero-mean spatial identity patterns. Dropping matched hard
negatives costs less ($0.812$), and a single donor is worse than mixing on both
privacy and structural retrieval. Learning the subspace is crucial: a
rank-equivalent random $U$ reaches only $0.912$, with reduced utility, so the
benefit is not due to $r$-channel perturbation alone.

\noindent\textbf{What the representation learned.}
Probes on frozen $E$, $G$ and $U$, with training heads removed, recover
identity from $\aid$ at $0.943$ but only at $0.712$ from $\zclin$, the expected
specialization. A frozen-encoder control shows the reverse ($0.781$ and
$0.869$), suggesting that joint learning creates this structure rather than
exploiting pre-existing information; the control is no fair privacy competitor,
as only a jointly learned encoder can react to $\Lc{adv}$. Two caveats remain:
identity is recoverable from $\zclin$ above chance, so the decomposition is
imperfect, and pathology is predictable from $\aid$ at $0.661$, so some clinical
information remains in the residual. With acquisition-matched negatives only,
linkability falls further to $0.742$ for LAPIS and $0.921$ for reconstruction,
so part of the attack is acquisition- rather than patient-based. The ranking holds
for all three attackers.

%% file: tables/ablation.tex
\begin{table}[t]
\centering
\caption{Ablations, with the attackers and protocol of Table~\ref{tab:main}.}
\label{tab:ablation}
\setlength{\tabcolsep}{3pt}
\small
\begin{tabular}{@{}l ccc@{}}
\hline
Variant & AUC$_{\mathrm{cross}}\downarrow$ & Rank-1$\downarrow$ & AUROC$\uparrow$ \\
\hline
\textbf{LAPIS (full)} & \textbf{0.791} & \textbf{34.7} & \textbf{0.786} \\
w/o replacement-aware training & 0.884 & 52.9 & 0.742 \\
Pooled identity adversary & 0.849 & 41.3 & 0.784 \\
w/o matched hard negatives & 0.812 & 36.8 & 0.779 \\
Direct donor residual & 0.848 & 47.2 & 0.773 \\
\hline
\end{tabular}
\end{table}

%% file: sections/7_conclusion.tex
\section{Conclusion}
\label{sec:conclusion}

LAPIS learns a compact patient-identity residual in the spatial latent
and trains the decoder to stay stable under its substitution. Against adapted
attackers on patient-disjoint cohorts, it lowers worst-case cross-study
linkability from $0.961$ to $0.791$ AUC at a $0.012$ macro AUROC cost relative
to reconstruction, inside the pre-set margin and at the lowest linkability of
any method clearing it; ablations show that residual learning and
replacement-aware decoding are both essential. Its weaknesses are specific: the
shared channel basis largely preserves anatomical geometry, so deformation
remains better against same-image identification, and identity is partly
recoverable from the complement, pointing to a geometry-sensitive residual as
the next direction.

%% file: sections/8_backmatter.tex
\section{Compliance with ethical standards}
\label{sec:ethics}

This study was conducted retrospectively on human subject data provided via open
access from the NIH Clinical Center (ChestX-ray14). The data are fully de-identified and publicly released, so no additional ethical approval was required; the dataset's terms of use were followed. No effort to
identify any individual aside from the statistical linkage measured here was
attempted, and no anonymized derivative images were distributed alongside this
work.

\section{Acknowledgments}
\label{sec:acknowledgments}

No funds were received for the study. There are no conflicts of interest to
report. ChatGPT (OpenAI) was used to assist with writing and Claude Opus~5.5
(Anthropic) to assist with coding; the authors reviewed and edited all
AI-assisted output and take full responsibility for the content of this paper.

%% file: references.bib
@inproceedings{wang2017chestxray,
  author    = {Xiaosong Wang and Yifan Peng and Le Lu and Zhiyong Lu and
               Mohammadhadi Bagheri and Ronald M. Summers},
  title     = {{ChestX-ray8}: Hospital-scale chest x-ray database and benchmarks on
               weakly-supervised classification and localization of common thorax
               diseases},
  booktitle = {Proceedings of the IEEE Conference on Computer Vision and Pattern
               Recognition (CVPR)},
  pages     = {2097--2106},
  year      = {2017}
}

@article{packhauser2022reid,
  author  = {Kai Packh{\"a}user and Sebastian G{\"u}ndel and Nicolas M{\"u}nster and
             Christopher Syben and Vincent Christlein and Andreas Maier},
  title   = {Deep learning-based patient re-identification is able to exploit the
             biometric nature of medical chest x-ray data},
  journal = {Scientific Reports},
  volume  = {12},
  number  = {1},
  pages   = {14851},
  year    = {2022}
}

@inproceedings{packhauser2023prichexy,
  author    = {Kai Packh{\"a}user and Sebastian G{\"u}ndel and Florian Thamm and
               Felix Denzinger and Andreas Maier},
  title     = {Deep learning-based anonymization of chest radiographs: A
               utility-preserving measure for patient privacy},
  booktitle = {Medical Image Computing and Computer Assisted Intervention (MICCAI)},
  pages     = {262--272},
  publisher = {Springer},
  year      = {2023}
}

@inproceedings{packhauser2023ldm,
  author    = {Kai Packh{\"a}user and Lukas Folle and Florian Thamm and Andreas Maier},
  title     = {Generation of anonymous chest radiographs using latent diffusion
               models for training thoracic abnormality classification systems},
  booktitle = {IEEE International Symposium on Biomedical Imaging (ISBI)},
  pages     = {1--5},
  year      = {2023}
}

@inproceedings{li2025gmia,
  author    = {Huiyu Li and Nicholas Ayache and Herv{\'e} Delingette},
  title     = {Generative medical image anonymization based on latent code projection
               and optimization},
  booktitle = {IEEE International Symposium on Biomedical Imaging (ISBI)},
  pages     = {1--4},
  year      = {2025}
}

@inproceedings{tian2025medreid,
  author    = {Yuan Tian and Kaiyuan Ji and Rongzhao Zhang and Yankai Jiang and
               Chunyi Li and Xiaosong Wang and Guangtao Zhai},
  title     = {Towards all-in-one medical image re-identification},
  booktitle = {Proceedings of the IEEE/CVF Conference on Computer Vision and Pattern
               Recognition (CVPR)},
  pages     = {30774--30786},
  year      = {2025}
}

@article{montenegro2024disentangle,
  author  = {Helena Montenegro and Jaime S. Cardoso},
  title   = {Anonymizing medical case-based explanations through disentanglement},
  journal = {Medical Image Analysis},
  volume  = {95},
  pages   = {103209},
  year    = {2024}
}

@inproceedings{ganin2015grl,
  author    = {Yaroslav Ganin and Victor Lempitsky},
  title     = {Unsupervised domain adaptation by backpropagation},
  booktitle = {Proceedings of the International Conference on Machine Learning (ICML)},
  pages     = {1180--1189},
  year      = {2015}
}

@inproceedings{khosla2020supcon,
  author    = {Prannay Khosla and Piotr Teterwak and Chen Wang and Aaron Sarna and
               Yonglong Tian and Phillip Isola and Aaron Maschinot and Ce Liu and
               Dilip Krishnan},
  title     = {Supervised contrastive learning},
  booktitle = {Advances in Neural Information Processing Systems (NeurIPS)},
  pages     = {18661--18673},
  year      = {2020}
}

@inproceedings{lezcano2019orthogonal,
  author    = {Mario Lezcano-Casado and David Mart{\'i}nez-Rubio},
  title     = {Cheap orthogonal constraints in neural networks: A simple
               parametrization of the orthogonal and unitary group},
  booktitle = {Proceedings of the International Conference on Machine Learning (ICML)},
  pages     = {3794--3803},
  year      = {2019}
}

@inproceedings{isola2017pix2pix,
  author    = {Phillip Isola and Jun-Yan Zhu and Tinghui Zhou and Alexei A. Efros},
  title     = {Image-to-image translation with conditional adversarial networks},
  booktitle = {Proceedings of the IEEE Conference on Computer Vision and Pattern
               Recognition (CVPR)},
  pages     = {1125--1134},
  year      = {2017}
}

@inproceedings{huang2017densenet,
  author    = {Gao Huang and Zhuang Liu and Laurens van der Maaten and
               Kilian Q. Weinberger},
  title     = {Densely connected convolutional networks},
  booktitle = {Proceedings of the IEEE Conference on Computer Vision and Pattern
               Recognition (CVPR)},
  pages     = {4700--4708},
  year      = {2017}
}

@inproceedings{liu2022convnext,
  author    = {Zhuang Liu and Hanzi Mao and Chao-Yuan Wu and Christoph Feichtenhofer
               and Trevor Darrell and Saining Xie},
  title     = {A {ConvNet} for the 2020s},
  booktitle = {Proceedings of the IEEE/CVF Conference on Computer Vision and Pattern
               Recognition (CVPR)},
  pages     = {11976--11986},
  year      = {2022}
}

@inproceedings{dosovitskiy2021vit,
  author    = {Alexey Dosovitskiy and Lucas Beyer and Alexander Kolesnikov and others},
  title     = {An image is worth 16x16 words: Transformers for image recognition at
               scale},
  booktitle = {International Conference on Learning Representations (ICLR)},
  year      = {2021}
}

@article{wang2004ssim,
  author  = {Zhou Wang and Alan C. Bovik and Hamid R. Sheikh and Eero P. Simoncelli},
  title   = {Image quality assessment: From error visibility to structural
             similarity},
  journal = {IEEE Transactions on Image Processing},
  volume  = {13},
  number  = {4},
  pages   = {600--612},
  year    = {2004}
}

@inproceedings{maximov2020ciagan,
  author    = {Maxim Maximov and Ismail Elezi and Laura Leal-Taix{\'e}},
  title     = {{CIAGAN}: Conditional identity anonymization generative adversarial
               networks},
  booktitle = {Proceedings of the IEEE/CVF Conference on Computer Vision and Pattern
               Recognition (CVPR)},
  pages     = {5446--5455},
  year      = {2020}
}
